# The effect of Hybrid Principal Components Analysis on the Signal Compression Functional Regression: With EEG-fMRI Application

Mohammad Fayaz[a],Alireza Abadi[b,c*] and Soheila Khodakarim [d]

[a] *PhD Student of Biostatistics, Department of Biostatistics, School of Allied Medical Sciences, Shahid Beheshti University of Medical Sciences, Tehran, Iran.*

[b] *Professor of Biostatistics, Department of Community Medicine, Faculty of medicine, Shahid Beheshti University of Medical Sciences, Tehran, Iran.*

[c] *Social Determinants of Health Research Center, Shahid Beheshti University of Medical Sciences, Tehran, Iran.*

[d]*Associate Professor, School of Allied Medical Sciences, School of Public Health and Safety, Shahid Beheshti University of Medical Sciences, Tehran, Iran*

* Corresponding Author: Dr. Alireza Abadi (alirezaabadi@gmail.com)

Professor of Biostatistics, Department of Community Medicine, Faculty of medicine, Shahid Beheshti University of Medical Sciences, Tehran, Iran. Social Determinants of Health Research Center, Shahid Beheshti University of Medical Sciences, Tehran, Iran.

Koodakyar Ave., Daneshju Blvd., Velenjak, Tehran, IRAN.

## 1- Introduction

The principal component regression (PCR) was used in the regression settings to tackle the collinearity problem between covariates [1-3]. With an introduction to the functional data [4], the functional extension of the Principal Component Analysis (FPCA) captured the underlying curves of the data and reduce their dimensionality [5]. We count different methods for estimating the number of FPCA such as fraction of variance explained (FVE), comparing the log-likelihood of different models with different numbers of FPCAs [6], Likelihood Ratio Test (LRT) [7], the mean square error (MSE), mean square prediction error (MSPE) [8] and Bayesian information criteria (BIC) [9]. The FPCA methods extend in different ways to deal with for example longitudinal dimension [10, 11], multilevel structure [12], multidimensional [13], and spatially dimension [7, 14-16]. The scalar-on-function (SOF), function-on-scalar (FOS), and function-on-function (FOF) regression and different supervised learning models use FPCA scores instead of the functional data.[4, 5, 17]. Some of them are adapted for predictions such as signal compression estimations [18, 19] for modeling the EEG and many regions of interest (ROI) of the Functional magnetic resonance imaging (fMRI) and Bayesian estimations [20, 21] with applications to the $\delta$-power of the electroencephalography (EEG) data and intraday physical activity with many covariates in the sleep study.

The hybrid PCA (HPCA) was introduced for situations that exists both scalar and functional data, called mixed or hybrid data, [4]. For example, the (EEG) data were analyzed with HPCA to extract the functional, longitudinal, and regional dimensions that they are frequency or time domain ERPs, the repeating stimuli, and the regions of the brain, respectively. The first two of them have functional characteristics and the last one is scalar. The HPCA extracts the subject-specific scores and eigencomponents of hybrid data [22]. Among the regression models for the hybrid data, we can count covariate-adjusted HPCA [23], the Semi-functional partial linear regression [24, 25], FOF regression with signal compression [18], and functional additive regression models [26]. In this research, we study the effects of HPCA decomposition of hybrid data on the prediction accuracy of the FOF regression with signal compression. We choose the number of HPCA based on the MSPE. Finally, we applied our methodology to the EEG-fMRI dataset. The paper has the following parts, *1-introduction* contains the problem definitions with some background, *2- The functional regression with HPCA* contains HPCA decomposition and FOF with multiple functional covariates, *3- The number of*

*HPCA* contains a methodology for choosing the best number of HPCA for prediction accuracy, *4- Simulations* contain two simulation scenarios, *5- Application to the EEG-fMRI dataset*, *6- Conclusion* contains the comparative study with some comments and *7- References*. We also provide supplementary materials including the R codes.

## 2- The Functional Regression with HPCA

### *2.1 – The HPCA Decomposition*

With the same notation and methodology of [22], we mention the HPCA decomposition briefly in this section. Assume that $Y_i(r,\omega,s)$ has three dimensions of the de-mean observation for the subject $i, \{i = 1, \dots, n\}$, the first dimension has scalar domain $r, r = 1, \dots, R$, the second dimension has functional domain $\omega, \omega \in \Omega$, and the third and desired dimension has functional domain $s, s \in \mathrm{S}$. We want to decompose $Y_i(r,\omega,s)$ with the HPCA to the eigencomponents and subject-specific scores. In this regard, we first estimate the covariance matrices. Because there are three dimensional matrices, we can estimate the eigencomponents of each dimension by marginal covariance and product them under the weak separability assumptions of covariances. In practice we have:

$$Y_i(r,\omega,s) = Z_i(r,\omega,s) + \epsilon_i(r,\omega,s)$$
$$\approx \sum_{k=1}^{K}\sum_{l=1}^{L}\sum_{m=1}^{M} \xi_{i,klm} V_k(r)\phi_l(\omega)\psi_m(s) + \epsilon_i(r,\omega,s) \qquad (1)$$

The $V_k(r)$ is the marginal eigenvector of the first dimension that has a scalar domain, $\phi_l(\omega)$ is the marginal eigenfunction of the second dimension that has a functional domain and $\psi_m(s)$ is the marginal eigenfunction of the three dimensions that has a functional domain. The $\xi_{i,klm}$ are the subject-specific scores obtained using the following inner product:

$$\xi_{i,klm} = < Z_i(r,\omega,s), V_k(r)\phi_l(\omega)\psi_m(s) > \; = \sum_{r=1}^{R} \iint Z_i(r,\omega,s) V_k(r)\phi_l(\omega)\psi_m(s) d\omega ds \qquad (2)$$

The $K, L$ and $M$ are the number of eigencomponents that marginally has at least 90% fraction of variation (FVE) of each dimension. Therefore, it is truncated with the maximum number of $KLM$ eigencomponents. The $\epsilon_i(r,\omega,s)$ are error terms.[22]

### *2.2 – The Function-on-multiple-Functions Regression*

In this part, we use the FOF regression [19], the response and covariates are functional data. For example Evoked Related Potential (ERP) curves on time or frequently domains. We first decompose hybrid data with HPCA and then reconstruct them with different number of HPCAs. The reconstructed hybrid data have three dimensions and we merge them to one dimension with a functional domain. Many applications work only with one dimension. For example, in the EEG dataset, $Y_i(r,\omega,s)$, we usually only interested in values of the ERP at the time domain, $s$. And the other two dimensions are the regional with the scalar domain, $r$ and the longitudinal dimension, $\omega$, with the functional domain.

Therefore, we first reconstruct the dataset in equation (3) with the specified number of the HPCA, $\hat{Y}_i(r,\omega,s)_{HPCA_KLM}$. We use hat notation because it is an estimation of the real observed value, $Y_i(r,\omega,s)$ and we subscript $HPCA_KLM$ to show the number of HPCAs, for example, $KLM$. In this step, we use the most variations of three dimensions (usually at least 90%) and we reconstruct the data based on them, therefor many unimportant variations were omitted. Second, we merge $\hat{Y}_i(r,\omega,s)_{HPCA_KLM}$ on the two other dimensions, for example, on the first, $r$, and second, $\omega$, dimesons. And as stated in the (4), we average point-wise on their domain of our desired dimension, for example, the third dimension, $s$.

$$\sum_{k=1}^{K}\sum_{l=1}^{L}\sum_{m=1}^{M}\xi_{i,klm}V_k(r)\phi_l(\omega)\psi_m(s)=\hat{Y}_i(r,\omega,s)_{HPCA_KLM} \qquad (3)$$

$$\frac{1}{R}\sum_{r=1}^{R}\int_{-\infty}^{+\infty}\hat{Y}_i(r,\omega,s)_{HPCA_KLM}d\,\omega=W_i(s)_{KLM} \qquad (4)$$

The advantage of this method instead of using only FPCA on the desired dimension, s, is that our method contains the most important variations of the other two-dimension in the dataset and it avoids considering the unimportant variations on the data and help to reduce the complexity of it. The functional covariate, $W_i(s)_{KLM}$, with the functional domain, $s$, and the number of HPCs, $KLM$ , is calculated by (4) for regression analysis in the next steps.

We assume that $W_i(s)_{KLM}$ (4) is a response function in the following FOF regression [19]:

$$W_i(s)_{KLM}=m(s)+\sum\nolimits_{j=1}^{p}\int_0^1\chi_j(g)\beta_j(g,s)dg+\varepsilon(s) \qquad (5)$$

The $\chi_1(g),\ldots,\chi_p(g)$ are predictive functions and $\beta_1(g,s),\ldots,\beta_p(g,s)$ are $p$ −coefficient functions, $m(s)$ and $\varepsilon(s)$ are intercept function and noise function. The focus of this model with signal compression methods is predicting the response [19].

**3- The number of HPCA**

The number of appropriate HPCA for $Y_i(r,\omega,s)$ in (1) is discussed at [22] with the estimating of the fraction of variance explained (FVE) in the mixed-effects models and the variance components estimations. In this part, we propose a method to select the number of HPCA based on the mean squared prediction error (MSPE) of the regression models. In (1) we have $KLM$ eigencomponents, but instead of using all of them in (3), we add each of them at each step:

$$\begin{gathered}\xi_{i,1}V_1(r)\phi_1(\omega)\psi_1(s)=\hat{Y}_i(r,\omega,s)_{HPCA_1}\\ \xi_{i,1}V_1(r)\phi_1(\omega)\psi_1(s)+\xi_{i,2}V_1(r)\phi_1(\omega)\psi_2(s)=\hat{Y}_i(r,\omega,s)_{HPCA_2}\\ \dots\\ \sum_{k=1}^{K}\sum_{l=1}^{L}\sum_{m=1}^{M}\xi_{i,klm}V_k(r)\phi_l(\omega)\psi_m(s)=\hat{Y}_i(r,\omega,s)_{HPCA_KLM}\end{gathered} \tag{6}$$

Therefore, $\{\hat{Y}_i(r,\omega,s)_{HPCA_q}, 1\le q\le KLM\}$ are estimations based on the first $q$ eigencomponents. We transform each of them with (4) to $\{W_i(s)_{HPCA_q}, 1\le q\le KLM\}$. For (5), we fit the regression model to all $KLM$ models with $W_i(s)_{HPCA_q}$ as a response curve and we predict $\{\widehat{W}_i(s)_{HPCA_q}, 1\le i\le n, 1\le q\le KLM\}$ (Algorithm-1)

| Algorithm 1 –FOF Regression (6) | |
|---|---|
| 1-The HPCA decomposition $Y_i(r,\omega,s)\approx\sum_{k=1}^{K}\sum_{l=1}^{L}\sum_{m=1}^{M}\xi_{i,klm}V_k(r)\phi_l(\omega)\psi_m(s)+\epsilon_i(r,\omega,s)$ | |
| 2- $for\ q=1\ to\ q=KLM$ | |
| 2-1 | The reconstruction based on the HPCA from 1 to $q$: |
| | $\sum_{k=1}^{K}\sum_{l=1}^{L}\sum_{m=1}^{M}\xi_{i,klm}V_k(r)\phi_l(\omega)\psi_m(s)=\hat{Y}_i(r,\omega,s)_{HPCA_q}$ |
| 2-2 | Merging on two dimensions: |
| | $\frac{1}{R}\sum_{r=1}^{R}\int_{-\infty}^{+\infty}\hat{Y}_i(r,\omega,s)_{HPCA_q}d\,\omega=W_i(s)_{HPCA_q}$ |
| 2-3 | Fit FOF based on the $W_i(s)_{HPCA_q}$ |
| | $W_i(s)_{HPCA_q}=m(s)+\sum_{j=1}^{p}\int_0^1\chi_j(g)\beta_j(g,s)dg+\varepsilon(s)$ |
| 2-4 | Predict the $\widehat{W}_i(s)_{HPCA_q}$ within (2-3) |

| | | |
|---|---|---|
| 2-5 | Calculate the MSPE | |
| | $Mean\ Squared\ Prediction\ Error =\ MSPE(q) = \frac{1}{n \times M} \sum_{i=1}^{n} \sum_{m=1}^{M} (W_i(s_m)_{HPCA_q} - \widehat{W}_i(s_m)_{HPCA_q})^2$<br>$\{i, 1 \leq i \leq n, number\ of\ subject\}, \{m, 1 \leq m \leq M, number\ of\ observation\ in\ each\ curve\}$ | |
| 3- Find the minimum $MSPE(q_{min}) =\ min\{q = 0\ or\ MSPE(q), 1 \leq q \leq KLM\}$. | | |
| 4- The best number of HPCA is $q_{min}$. | | |

This methodology can be applied to any other regression problem which deals with the hybrid data.

## 4- Simulations

### *4-1- Simulation Scenario 1*

*The function-on-function regression*: The data simulation process has two steps: We first generate the $Y_i(r, \omega, s)$ with two sample size,1) $\{i, i = 1, \ldots, 20\}$ and 2) $\{i, i = 1, \ldots, 140\}$ and two status for the 1) complete and 2) sparse observation for $\omega$ as mentioned in the [22], then we merge each of them with (1) to (5), as $W_i(s)_{HPCA_0}$.
In the second step, we generate the fof regression data as mentioned at [19] with 10 covariates functions, $\theta(g, s)$, and two status for coefficients 1) all 10 coefficients, $\theta(g, s)$, have a value greater than 0 and 2) only 50% of $\theta(g, s)$ have the non-zero value. Therefore, we have $A_i(s) = m(s) +\ \sum_{j=1}^{10} \int_0^1 \chi_j(g)\theta_j(g, s)dg + \varepsilon(s)$ .
We combine the first and second steps with $W_i(s)_{HPCA_0} + A_i(s) = D_i(s)_{HPCA_0}$ and we fit the fof regression with the following equation, as (6):

$$D_i(s)_{HPCA_0} = m(s) + \sum_{j=1}^{p} \int_0^1 \chi_j(g)\beta_j(g, s)dg + \varepsilon(s)$$

In each model, we split the data to the train (70%) and test (30%) and we fit the model on the training dataset. We calculate the following summary statistics for the coefficients:

MSE $\beta_1(s, t) = \frac{1}{n \times S \times T} \sum_{i=1}^{n} \sum_{s=1}^{S} \sum_{t=1}^{T} (\beta_1(s, t) - \hat{\beta}_1(s, t))^2$

….

MSE $\beta_{10}(s, t) = \frac{1}{n \times S \times T} \sum_{i=1}^{n} \sum_{s=1}^{S} \sum_{t=1}^{T} (\beta_{10}(s, t) - \hat{\beta}_{10}(s, t))^2$

For the prediction accuracy of model in the train and test, separately:

$$MSPE_{Pred(train)} = \frac{1}{n_{train}} \sum_{i=1}^{n_{train}} \sum_{j=1}^{J} (D_{i,train}(s_j)_{HPCA_0} - \widehat{D}_{i,train}(s_j)_{HPCA_0})^2$$

$$MSPE_{Pred(test)} = \frac{1}{n_{test}} \sum_{i=1}^{n_{test}} \sum_{j=1}^{J} (D_{i,test}(s_j)_{HPCA_0} - \widehat{D}_{i,test}(s_j)_{HPCA_0})^2$$

$$COR_{Pred(train)} = COR(D_{i,train}(s)_{HPCA_0}\ ,\widehat{D}_{i,train}(s)_{HPCA_0})$$

$$COR_{Pred(test)} = COR(D_{i,test}(s)_{HPCA_0}\ ,\widehat{D}_{i,test}(s)_{HPCA_0})$$

We also compare the elapsed, system, and user time for computations. We repeat the iteration 100 times. The median (the first quartile, the third quartile) in all simulations were reported.

We explain briefly the result: in the table 1, the $MSE\ \beta_{1,\ldots,10}(s,t)$ are greater for the complete rather than 50 % sparcity of $\boldsymbol{\beta}$. The $MSPE_{Pred(train)}$ are not very different in all situations with each other, but $MSPE_{Pred(test)}$ are different. The first difference is the $MSPE_{Pred(test)}$ for complete $\boldsymbol{\beta}$ are greater than the 50 % sparsity of $\boldsymbol{\beta}$. The second difference is with increasing the sample size, their values are smaller than big sample size. But its value for the sparsity for $\omega$ are not different with other. The $COR_{Pred(train)}$ are negligibly bigger than $COR_{Pred(test)}$ , but they are not very different between scenarios. The elapsed time for 50 % sparsity of $\boldsymbol{\beta}$ are greater than the complete $\boldsymbol{\beta}$ and they increase significantly with the number of sample size. But this behaviour are not observed for the system time. Therefore, the sparsity of $\boldsymbol{\beta}$ will affect the model performance and the sparsity of the $\boldsymbol{\omega}$ don't have any significant effect on them.

### *4-2- Simulation Scenario 2*

This simulation scenario is very similar to the simulation scenario 1, but in this one we study the number of HPCA. Again, we first generate the $Y_i(r,\omega,s)$ $\{i, i = 1, \ldots, 20\}$ as mentioned in the [22], then we merge them with (4) in three ways: 1) Without HPCA decomposition, as $W_i(s)_{HPCA_0}$ , 2) With only the first HPCA decomposition ,as $W_i(s)_{HPCA_1}$, 3) With All HPCA decomposing, as $W_i(s)_{HPCA_All}$ (All HPCA has 90% FVE on each dimension).We also generate the fof regression data as mentioned at [19] with 10 covariates functions , $\theta(g,s)$, and two status for coefficients 1) all 10 coefficients , $\theta(g,s)$, have the value greater than 0 and 2) only 50% of $\theta(g,s)$ have the non-zero value. Therefore, we have $A_i(s) = m(s) + \sum_{j=1}^{10} \int_0^1 \chi_j(g)\theta_j(g,s)\, dg + \varepsilon(s)$ . We define:

$$W_i(s)_{HPCA_0} + A_i(s) = D_i(s)_{HPCA_0}$$

$$D_i(s)_{HPCA_0} = m(s) + \sum_{j=1}^{p} \int_0^1 \chi_j(g)\beta_j(g,s)dg + \varepsilon(s)$$

$$W_i(s)_{HPCA_1} + A_i(s) = D_i(s)_{HPCA_1}$$

$$D_i(s)_{HPCA_1} = m(s) + \sum_{j=1}^{p} \int_0^1 \chi_j(g)\beta_j(g,s)dg + \varepsilon(s) \qquad (7)$$

$$W_i(s)_{HPCA_All} + A_i(s) = D_i(s)_{HPCA_All}$$

$$D_i(s)_{HPCA_All} = m(s) + \sum_{j=1}^{p} \int_0^1 \chi_j(g)\beta_j(g,s)dg + \varepsilon(s)$$

In each model, $D_i(s)_{HPCA_0}, D_i(s)_{HPCA_1}$ *and* $D_i(s)_{HPCA_All}$,we split the data to the train (70%) and test (30%) and we fit the models on the train dataset. We calculate the summary statistics as mentioned in the scenario 2. We repeat the iteration 100 times. The median (the first quartile, the third quartile) in all simulations were reported.

We explain briefly the result: in the table 2, the most of $MSE\ \beta_{1,\dots,10}(s,t)$ are greater for the complete rather than 50 % sparcity of $\boldsymbol{\beta}$. The $MSPE_{Pred(train)}$ are 0.1001 (0.084, 0.1294) for without HPCA, 0.0028 (0.0001,0.0176) with the first HPCA and 0.0174 (0.0024,0.0409) for the all HPCA in the complete $\boldsymbol{\beta}$ settings. It stated that the first HPCA has the lowest MSPE among others (2% and 16% of wihout HPCA and with all HPCA, respectively). The same pattern exists for the $MSPE_{Pred(test)}$. The $COR_{Pred(train)}$ and $COR_{Pred(test)}$ are high (more than 90%) and the differences are negligible. The elapsed time are lowest for the first HPCA model among others.

## 5- Application to the EEG-fMRI dataset

In this application, we use the EEG-fMRI dataset which was analysed with FOF regression previously[19]. The novelty of this new analysis are: 1) We analyse the event-related potential (ERP) with the time instead of frequency domain as response function and 2) we also study the effect of HPCA on the prediction accuracy of FOF regression. The dataset is related to an experiment to study the internal attentions in the brain with 17 participants (6 females, mean age is 27.7 years) [28]. There are two tasks for each experiment including auditory and visual stimuli. And there are two stimuli in each task, standard and target. The auditory task includes standard stimuli (390 Hz tone) and target stimuli (laser gun sound). The visual task includes standard stimuli (small green circle on isoluminant grey background, 1.5-degree visual angle) and target stimuli (large red circle on isoluminant grey background, 3.45-degree visual angle). The dataset was downloaded from openneuro.org. The EEG was captured based on 64 channels and the preprocessing steps are artifact removal, baseline correction and epoching (-200 to 800 milliseconds of each stimuli) that they were done with EEGLAB[29] . The fMRI dataset was captured with 3 Tesla Philips scanner and the dataset was preprocessed through slice-timing correction (SCT), normalization, coregistering , smoothing and region of interest (ROI) extraction with SPM 12 toolbox (http://www.fil.ion.ucl.ac.uk/spm/). The ERP waves are response functions and the ROI time series for different Brodmann Areas in the brain are functional covariates. We conduct two analyses: the first one is choosing the best number of HPCA and the second is FOF regression for EEG-fMRI by the region of the brain.

We first decompose the ERP curves with the HPCA from first to the total number of HPCAs for the visual standard task. The KML is 14 that captures 90% of FVE in each dimension. Then we fit model (7) on the training samples and compare with the test. We do this procedure 100 times with different training/testing samples. Finally, we compare the $MSPE_{Pred(train)}$, $MSPE_{Pred(test)}$, $COR_{Pred(train)}$, $COR_{Pred(test)}$ and $Time_{elapsed}$ for different number of HPCAs.

We see the first HPCA has the lowest MSPE in both train/test. With increasing number of HPCA, the MSPE increases. Notably, the model without HPCA has the largest MSPE. (figure 1) The same pattern are seen in the correlation and the elapsed time. (figure 2 and 3)

Therefore, we use the first HPCA and analyses the whole dataset. In this regard, we analyse the whole brain and by the regions and compare the MSPE and Correlation in the train/test for the auditory and visual tasks. The MSPE for the auditory in all brain regions for standard and target stimulis are 0.00007 and 0.01337 and for visually are 0.00235 and 0.41549. The result are also provided by the different regions of the brain. (table 3)

All computations are done with R x64 4.0.2 and RStudio version Version 1.2.5042 in the server with 24 GB RAM, 64-bit Operating System, Intel® Xeon® CPU, X5670 @ 2.93 GHz 2.93 GHz (2 processors) from Turin Cloud Services (turin.ipm.ir). All R codes with their instructions are attached in the supplementary files.

## 6- Conclusion

We propose a method for using the HPCA in the functional regressions and choosing the best number of HPCA. In this regard, we first decompose the hybrid data with HPCA, then we reconstruct the observation based on the HPCA eigencompoents. This method helps us to reconstruct the data with the most important eigencomponents of each dimension and eliminates the noise and other negligible variations. Note that the various functional regressions based on the type of estimations, responses, and covariates are studied, and we use signal-compression fof which were previously published with the EEG-fMRI dataset. [19]

Note that this functional regression works with one-dimensional function, for example, it is common to use the ERP over time or frequency in the EEG dataset. We propose a method to convert the hybrid data to one-dimensional data by pooling and averaging two other dimensions as mentioned in the functional regression models. Note that instead of using the observed hybrid data, we use the reconstructed data from the HPCA eigencomponents. We also find the best number of HPCA based on the functional regression MSPE. Surprisingly, the simulations and real-world application with the EEG-fMRI show that the best performance of the regression was achieved by only the first eigencomponents. We can count the following reasons to explain this situation, first, the simulations are designed to show the EEG structure of the data, which has three dimensions, and most of the variations from two other dimesons are captured with the first HPCA; second, the higher number of HPCA were used in the reconstructed data, it is getting closer to the observed data, therefore their complexity increases that this matter affects the prediction accuracy in the functional regression. The future direction of the proposed method is using other indices than MSPE of prediction to choose the number of HPCA, using different types of functional regressions, and applying the methodology on their applications other than the EEG structure. Finally, we conclude that our methodology improves the prediction of the experiments with the EEG datasets. And we recommend that instead of using the functional PCA on the desired dimension, for example, ERP over time or frequency, reconstruct the data with HPCA and average it on the other two dimensions for functional regression models.

## 7- References:

Table 1 – Percentiles 50% (25%, 75%) of the simulation scenario 1.

| | ***Number of Subject*** | | | | | | | |
|---|---|---|---|---|---|---|---|---|
| | $\boldsymbol{n=20}$ | | | | $\boldsymbol{n=140}$ | | | |
| $\omega$ | $Complete$ | | $Sparcity$ | | $Complete$ | | $Sparcity$ | |
| $\boldsymbol{\beta}$ | $Complete$ | $50\ \%\ Sparcity$ | $Complete$ | $50\ \%\ Sparcity$ | $Complete$ | $50\ \%\ Sparcity$ | $Complete$ | $50\ \%\ Sparcity$ |
| MSE $\beta_1(s,t)$ | 20.61 (9.94,33.4) | 1.96 (0.03,22.18) | 17.61 (8.59,34.25) | 5.02 (0.08,19.23) | 29.39 (18.05,46.26) | 0.0132 (0,22.4816) | 33.29 (19.93,48.03) | 8.83 (0.003,32.637) |
| MSE $\beta_2(s,t)$ | 1.5 (1.03,2.89) | 0.5 (0.1,1.08) | 1.47 (0.88,2.42) | 0.58 (0.1,1.32) | 0.18 (0.14,0.23) | 0.0895 (0.0032,0.1596) | 0.19 (0.15,0.22) | 0.098 (0.001,0.186) |
| MSE $\beta_3(s,t)$ | 2.75 (1.73,4.66) | 0.99 (0.08,2.82) | 2.86 (1.93,4.92) | 1.14 (0.06,2.4) | 1.46 (1.04,2.32) | 0.018 (0.0001,1.1931) | 1.59 (1.02,2.42) | 0.042 (0.003,1.148) |
| MSE $\beta_4(s,t)$ | 2.96 (1.86,5.29) | 1.32 (0.13,2.85) | 3.55 (1.95,5.54) | 0.76 (0.04,2.59) | 2.4 (1.71,3.7) | 0.9192 (0.006,2.2836) | 2.76 (1.83,4.06) | 0.047 (0.003,2.845) |
| MSE $\beta_5(s,t)$ | 4.03 (2.83,6.35) | 1.23 (0.07,3.88) | 3.3 (2.11,5.14) | 1.15 (0.1,3.47) | 2.94 (2.13,4.72) | 0.8649 (0.0005,3.3766) | 3.15 (2.01,4.3) | 0.654 (0.009,3.227) |
| MSE $\beta_6(s,t)$ | 3.93 (2.43,6.27) | 0.57 (0.05,4.31) | 3.7 (2.17,5.03) | 1.17 (0.11,3.18) | 3.33 (1.84,5.19) | 1.1584 (0.0052,3.4304) | 3.29 (2.2,4.68) | 1.436 (0.004,4.286) |
| MSE $\beta_7(s,t)$ | 3.73 (2.36,5.79) | 0.45 (0.07,3.06) | 3.93 (2.29,5.84) | 1.09 (0.1,3.92) | 2.99 (1.82,4.48) | 1.0422 (0.0028,3.8262) | 2.74 (1.87,5.23) | 1.809 (0.012,3.895) |
| MSE $\beta_8(s,t)$ | 3.56 (2.04,5.43) | 0.91 (0.08,3.9) | 3.91 (2.61,6.84) | 0.56 (0.03,3.47) | 3.41 (1.94,5.47) | 0.0603 (0.0025,3.5484) | 4.02 (2,5.93) | 0.037 (0.001,2.989) |
| MSE $\beta_9(s,t)$ | 4.2 (2.76,6.67) | 0.4 (0.03,3.55) | 4.02 (2.44,5.95) | 1.07 (0.07,3.53) | 3.41 (2.29,5.51) | 1.7362 (0.0086,3.9769) | 4.1 (2.14,6.39) | 0.051 (0.003,3.411) |
| MSE $\beta_{10}(s,t)$ | 3.4 (2.1,6.17) | 2.05 (0.1,5.22) | 3.72 (2.12,6.06) | 1.13 (0.05,3.74) | 4.1 (2.52,6.52) | 0.0398 (0.001,3.2849) | 3.75 (2.32,6.28) | 0.671 (0.004,4.251) |
| $MSPE_{Pred(train)}$ | 0.09 (0.07,0.12) | 0.09 (0.08,0.11) | 0.09 (0.07,0.12) | 0.09 (0.07,0.1) | 0.09 (0.09,0.1) | 0.0975 (0.0933,0.1012) | 0.1 (0.09,0.1) | 0.098 (0.096,0.102) |
| $MSPE_{Pred(test)}$ | 0.37 (0.25,0.5) | 0.19 (0.14,0.32) | 0.33 (0.25,0.54) | 0.21 (0.15,0.33) | 0.12 (0.11,0.13) | 0.1101 (0.1045,0.1182) | 0.12 (0.11,0.13) | 0.114 (0.106,0.12) |
| $COR_{Pred(train)}$ | 0.98 (0.98,0.99) | 0.97 (0.96,0.98) | 0.98 (0.98,0.99) | 0.97 (0.96,0.98) | 0.98 (0.98,0.99) | 0.9613 (0.9559,0.9784) | 0.98 (0.98,0.99) | 0.97 (0.957,0.983) |
| $COR_{Pred(test)}$ | 0.94 (0.9,0.96) | 0.94 (0.92,0.96) | 0.94 (0.9,0.97) | 0.94 (0.91,0.96) | 0.98 (0.97,0.98) | 0.9564 (0.9511,0.971) | 0.98 (0.97,0.98) | 0.964 (0.952,0.979) |
| $Time_{user}$ | 48.33 (35.55,70.02) | 101.98 (55.02,133.85) | 51.64 (37.22,74.64) | 102.54 (53.44,136.38) | 95.5 (79.82,120.1) | 322.955 (105.965,381.245) | 97.94 (79.59,123.5) | 169.76 (88.53,385.28) |
| $Time_{system}$ | 0.07 (0.05,0.09) | 0.09 (0.07,0.13) | 0.06 (0.04,0.09) | 0.07 (0.05,0.11) | 0.24 (0.18,0.3) | 0.415 (0.32,0.51) | 0.11 (0.07,0.14) | 0.11 (0.08,0.15) |
| $Time_{elapsed}$ | 48.4 (35.59,70.1) | 102.11 (55.1,133.98) | 51.66 (37.26,74.71) | 102.69 (53.56,136.51) | 95.94 (80.28,120.36) | 323.425 (106.585,381.8525) | 98.27 (79.93,123.84) | 170.01 (88.78,385.73) |

Table 2 – Percentiles 50% (25%, 75%) of the simulation scenario 2.

| HPCA | NO HPCA | | First HPCA | | All HPCA | |
|---|---|---|---|---|---|---|
| $\boldsymbol{\beta}$ | $Complete$ | $50\ \%\ Sparcity$ | $Complete$ | $50\ \%\ Sparcity$ | $Complete*$ | $50\ \%\ Sparcity\ **$ |
| MSE $\beta_1(s,t)$ | 14.5112 (8.3463,27.4955) | 19.1862 (10.6082,27.8363) | 14.6201 (7.2417,25.9006) | 16.317 (10.498,27.322) | 15.925 (8.7129,30.6679) | 19.0602 (11.577,28.7054) |
| MSE $\beta_2(s,t)$ | 1.4757 (0.8491,2.6039) | 0.6611 (0.0946,1.3561) | 3.0187 (1.6888,4.1153) | 0.94 (0.042,2.369) | 2.0448 (1.0246,2.8579) | 0.7309 (0.0696,1.5715) |
| MSE $\beta_3(s,t)$ | 2.6643 (1.6905,4.3293) | 0.783 (0.0967,2.6223) | 3.4721 (2.1247,5.1566) | 0.873 (0.026,3.042) | 2.8849 (1.6589,4.6093) | 0.8535 (0.1357,2.6239) |
| MSE $\beta_4(s,t)$ | 3.0949 (2.0827,6.0484) | 0.4074 (0.0523,2.5258) | 3.2804 (2.2413,5.7897) | 0.134 (0.003,2.313) | 3.0776 (2.0219,5.9766) | 0.4004 (0.0257,2.015) |
| MSE $\beta_5(s,t)$ | 3.9203 (2.3426,5.7395) | 1.0133 (0.1678,3.2986) | 4.0012 (2.4556,6.741) | 0.439 (0.094,3.171) | 3.9086 (2.6642,6.2414) | 0.9969 (0.129,3.2651) |
| MSE $\beta_6(s,t)$ | 4.221 (2.3879,6.6958) | 1.4369 (0.122,4.3104) | 3.8636 (2.4522,6.0266) | 1.366 (0.085,4.27) | 3.7442 (2.4925,6.9392) | 1.592 (0.1351,4.2872) |
| MSE $\beta_7(s,t)$ | 3.3584 (2.2642,5.8811) | 0.426 (0.0373,2.8341) | 3.3409 (2.1588,5.1684) | 0.385 (0.007,3.052) | 3.3756 (2.2225,4.9557) | 0.5478 (0.0369,2.944) |
| MSE $\beta_8(s,t)$ | 3.8073 (2.389,5.5397) | 0.7336 (0.083,3.8915) | 3.6125 (2.5294,5.5124) | 0.423 (0.018,3.47) | 3.7223 (2.4792,5.6815) | 0.962 (0.0842,3.6444) |
| MSE $\beta_9(s,t)$ | 3.48 (2.471,5.5648) | 0.4391 (0.063,2.4887) | 3.4562 (2.3211,5.3077) | 0.236 (0.007,2.724) | 3.5074 (2.3809,5.748) | 0.4747 (0.1196,2.7758) |
| MSE $\beta_{10}(s,t)$ | 3.7703 (2.2949,5.2103) | 1.1676 (0.1598,3.8849) | 3.5595 (2.3031,5.6235) | 1.239 (0.034,3.666) | 3.7815 (2.3975,5.914) | 1.3054 (0.1028,4.0518) |
| $MSPE_{Pred(train)}$ | 0.1001 (0.084,0.1294) | 0.0953 (0.0753,0.1118) | 0.0028 (0.0001,0.0176) | 0.004 (0.001,0.009) | 0.0174 (0.0024,0.0409) | 0.0109 (0.0037,0.029) |
| $MSPE_{Pred(test)}$ | 0.3207 (0.2452,0.5342) | 0.2967 (0.2107,0.4604) | 0.1118 (0.0439,0.2274) | 0.087 (0.044,0.27) | 0.2484 (0.1498,0.4314) | 0.2529 (0.1223,0.4378) |
| $COR_{Pred(train)}$ | 0.9801 (0.9691,0.9873) | 0.9829 (0.9755,0.9884) | 0.9994 (0.9962,1) | 0.999 (0.998,1) | 0.9971 (0.9922,0.9996) | 0.9979 (0.9942,0.9994) |
| $COR_{Pred(test)}$ | 0.9334 (0.9005,0.9647) | 0.942 (0.915,0.9684) | 0.9801 (0.9502,0.9928) | 0.982 (0.958,0.993) | 0.962 (0.9265,0.9765) | 0.955 (0.9277,0.9792) |
| $Time_{user}$ | 58.095 (42.135,83.1375) | 55.82 (39.2425,82.46) | 15.44 (14.305,17.685) | 14.89 (12.83,17.863) | 34.72 (30.61,41.42) | 33.67 (28.37,40.185) |
| $Time_{system}$ | 0.05 (0.03,0.08) | 0.05 (0.03,0.06) | 0.05 (0.03,0.0825) | 0.04 (0.02,0.11) | 0.08 (0.05,0.11) | 0.05 (0.03,0.08) |
| $Time_{elapsed}$ | 58.14 (42.24,83.2425) | 55.87 (39.2775,82.515) | 15.61 (14.4425,17.815) | 14.935 (12.868,18.265) | 34.89 (30.765,41.7125) | 33.75 (28.54,40.27) |

Number of iteration 100 times, * has 96 iteration and ** has 99 iterations.

Table 3 – The FOF result from EEG-fMRI Dataset by the region of the brain.

| Tasks | Region | Stimuli | MSPE | | Correlation | |
|---|---|---|---|---|---|---|
| | | | Train | Test | Train | Test |
| Auditory | All | Standard | 0.00007 | 0.07061 | 0.99997 | 0.96444 |
| | | Target | 0.01337 | 0.16030 | 0.99329 | 0.91817 |
| | Frontal | Standard | 0.00008 | 0.12564 | 0.99996 | 0.93640 |
| | | Target | 0.00174 | 0.19295 | 0.99913 | 0.90111 |
| | Central | Standard | 0.00004 | 0.04605 | 0.99998 | 0.97686 |
| | | Target | 0.00177 | 0.23528 | 0.99911 | 0.87700 |
| | Occipital | Standard | 0.00014 | 0.00123 | 0.99993 | 0.99939 |
| | | Target | 0.00024 | 0.00178 | 0.99988 | 0.99911 |
| | Parietal | Standard | 0.00018 | 0.03787 | 0.99991 | 0.98092 |
| | | Target | 0.02034 | 0.15549 | 0.98977 | 0.92011 |
| | Left Temporal | Standard | 0.00116 | 0.02427 | 0.99942 | 0.98784 |
| | | Target | 0.00005 | 0.00062 | 0.99998 | 0.99969 |
| | Right Temporal | Standard | 0.00040 | 0.06260 | 0.99980 | 0.96833 |
| | | Target | 0.02455 | 0.12669 | 0.98763 | 0.93528 |
| Visually | All | Standard | 0.00235 | 0.14560 | 0.99882 | 0.92580 |
| | | Target | 0.41549 | 0.52133 | 0.79498 | 0.72420 |
| | Frontal | Standard | 0.01259 | 0.08244 | 0.99368 | 0.95795 |
| | | Target | 0.06498 | 0.17451 | 0.96734 | 0.91379 |
| | Central | Standard | 0.00009 | 0.09191 | 0.99995 | 0.95308 |
| | | Target | 0.53342 | 0.64774 | 0.71244 | 0.61246 |
| | Occipital | Standard | 0.00015 | 0.00149 | 0.99993 | 0.99926 |
| | | Target | 0.00041 | 0.00186 | 0.99980 | 0.99907 |
| | Parietal | Standard | 0.00008 | 0.06213 | 0.99996 | 0.96845 |
| | | Target | 0.32912 | 0.39895 | 0.83103 | 0.78537 |
| | Left Temporal | Standard | 0.00460 | 0.04253 | 0.99769 | 0.97848 |
| | | Target | 0.02002 | 0.07749 | 0.98993 | 0.96081 |
| | Right Temporal | Standard | 0.00132 | 0.05436 | 0.99934 | 0.97249 |
| | | Target | 0.11777 | 0.23887 | 0.93919 | 0.88745 |

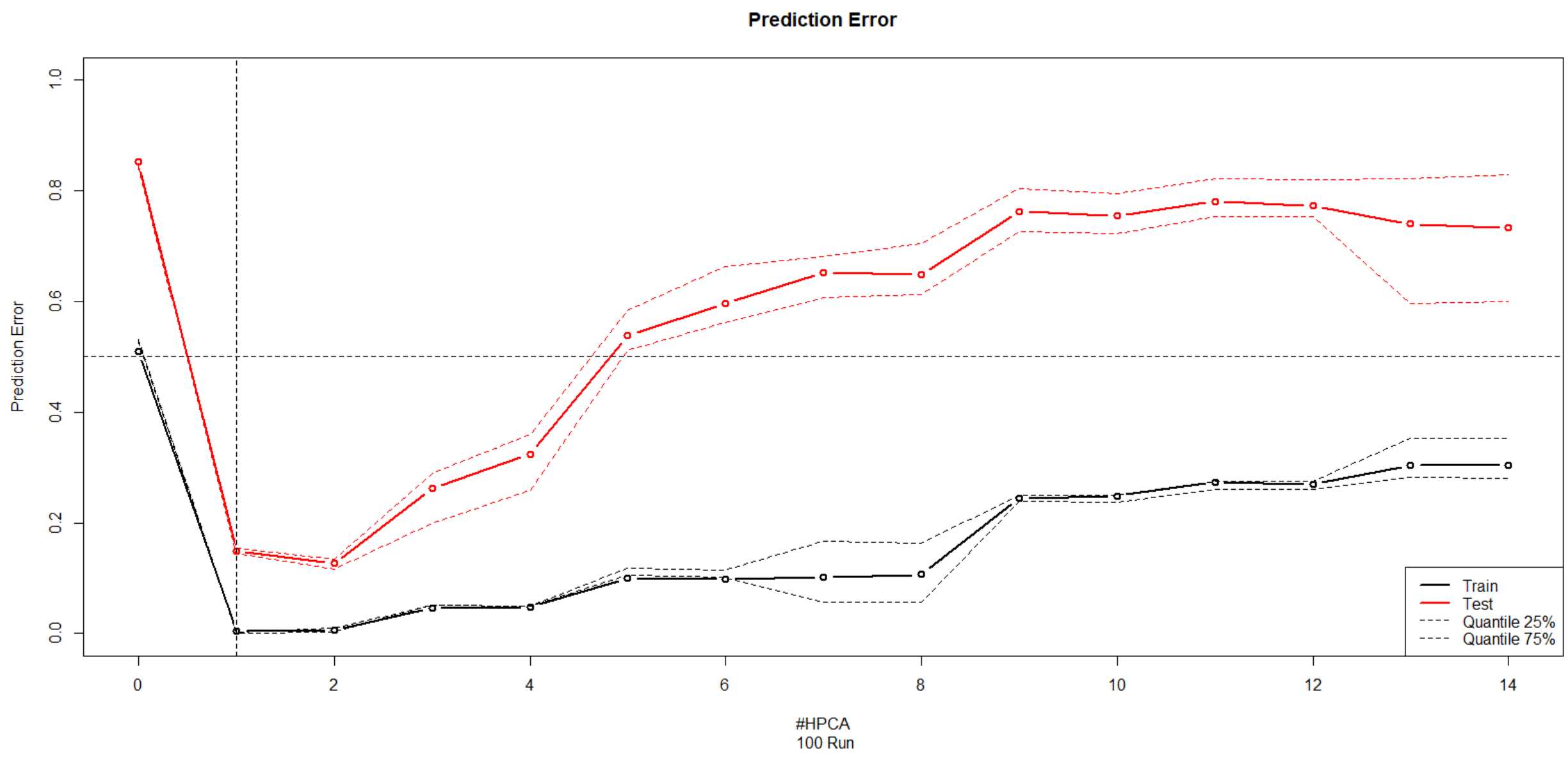

Figure 1- The MSPE prediction error by the number of HPCA in training/test samples.

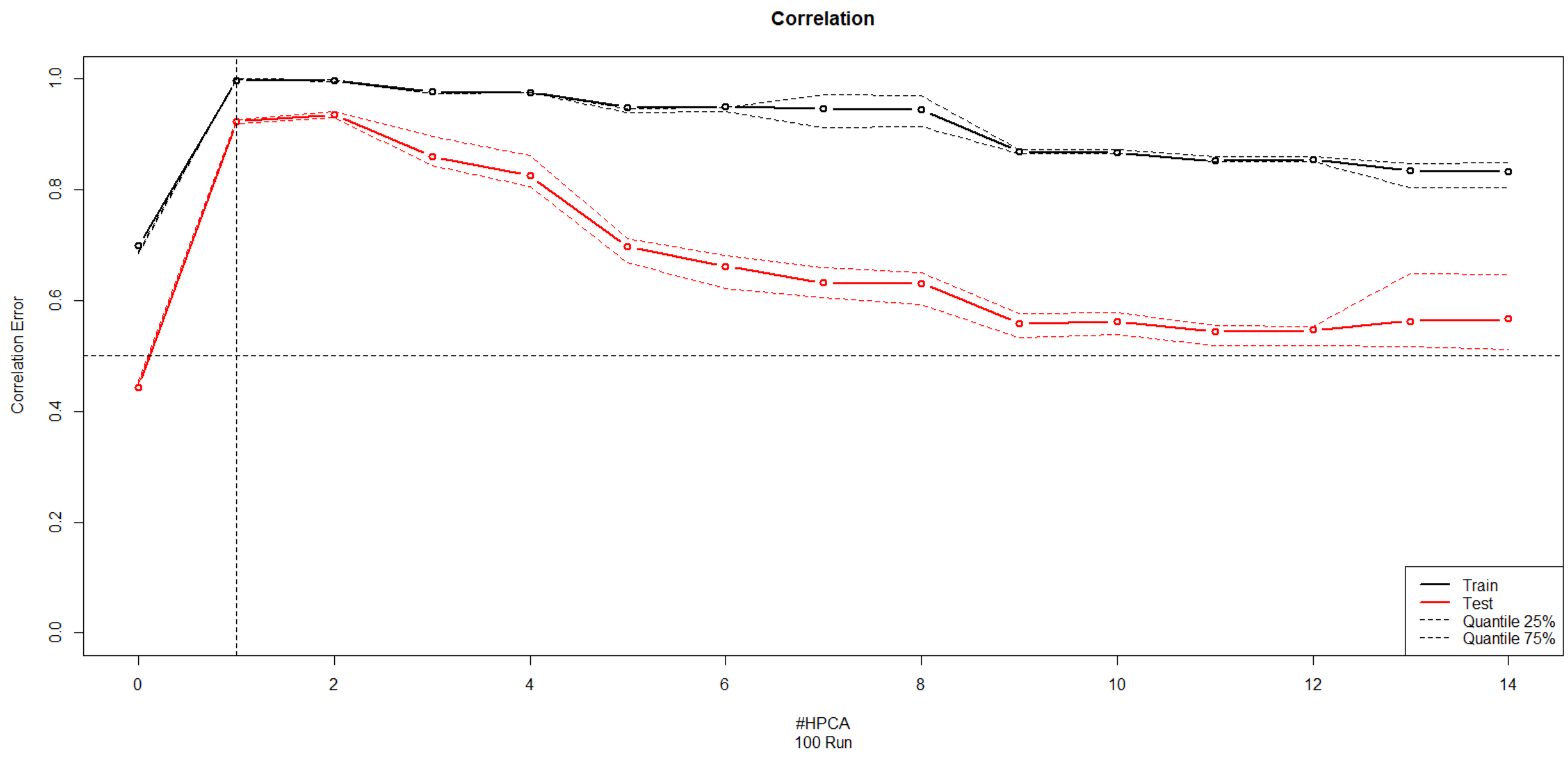

Figure 2- The Correlation by the number of HPCA in training/test samples.

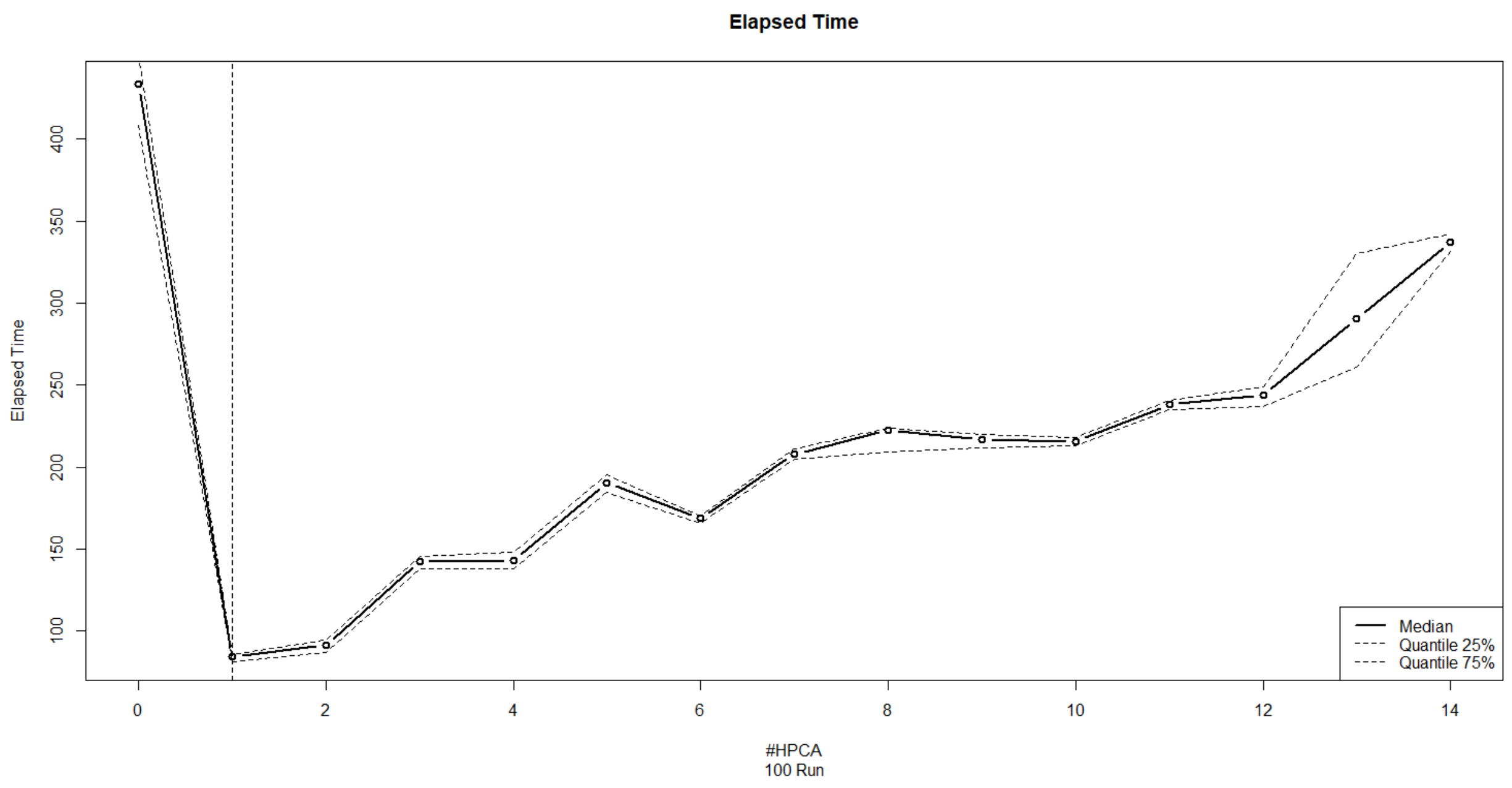

Figure 3- The elapsed time by the number of HPCA in training/test samples.